\documentclass[twocolumn,pra,superscriptaddress]{revtex4-1}

\usepackage{epsfig}
\usepackage{graphicx}
\usepackage{amsmath}
\usepackage{amssymb}
\usepackage{epstopdf}
\usepackage{xcolor}
\usepackage{subfigure}
\usepackage{bm}
\usepackage{dsfont}
\usepackage{hyperref}
\usepackage{braket}
\usepackage{todonotes}
\usepackage{physics}
\usepackage{comment}
\usepackage{marginnote}
\usepackage[normalem]{ulem}
\usepackage{mathtools}
\newcommand{\mi}{\mathrm{i}}

\begin{document}

\title{Intra-atomic frequency comb based photonic quantum memory using single-atom-cavity setup }

\author{Chanchal}
\affiliation{Department of Physical Sciences, Indian Institute of Science Education and Research, Mohali, Punjab 140306, India } 

\author{G.~P.~Teja}
\affiliation{Department of Physical Sciences, Indian Institute of Science Education and Research, Mohali, Punjab 140306, India }

\author{Sandeep K.~Goyal}
\email{skgoyal@iisermohali.ac.in}
\affiliation{Department of Physical Sciences, Indian Institute of Science Education and Research, Mohali, Punjab 140306, India }

\begin{abstract}
  On-demand and efficient storage of photons is an essential element in quantum information processing and long-distance quantum communication. Most of the quantum memory protocols require bulk systems in order to store photons. However, with the advent of integrated photonic chip platforms for quantum information processing, on-chip quantum memories are highly sought after. In this paper, we propose a protocol for multi-mode photonic quantum memory using only single-atom-cavity setup. We show that  a single atom containing a frequency comb coupled to an optical cavity can store photons efficiently. Further, this scheme can also be used to store polarization states of light. As examples, we show that the Rubidium and Cesium atoms coupled to nanophotonic waveguide cavities can serve as promising candidates to realize our scheme. This provides a possibility of a robust and efficient on-chip quantum memory to be used in integrated photonic chips.
 
\end{abstract}

\maketitle

\section{Introduction}

A photonic quantum memory is a device that can store and re-emit photons on demand~\cite{lvovsky2009,simon2010,heshami2016}. It is an essential component in quantum information processing applications such as quantum networks~\cite{kimble2008,simon2017}, quantum repeaters~\cite{qrepeater2007,qrepeater2011} and long range quantum communication~\cite{duan2001}. In a typical atomic ensemble based quantum memory, a weak light pulse is absorbed as delocalized atomic excitation over all the atoms in the ensemble. This collective atomic excitation is then transferred to a long lived spin state of the atoms using control pulses. In order to retrieve the photons from the atomic ensemble, a trigger pulse is used to transfer the excitation from the long-lived spin state to the excited state of the atom, which emits the photons at a desired time~\cite{simon2010,heshami2016}.

Some of the commonly used quantum memory protocols are electromagnetically induced transparency (EIT)~\cite{fleischhauer2002,david2010,hsiao2018,wang19}, controlled reversible inhomogeneous broadening (CRIB)~\cite{moiseev2001,kraus2005,alexander2006,sangouard2007},
gradient echo memory (GEM),~\cite{hetet08,hedges2010,hosseini11}, 
Raman memory~\cite{klein09,kozhekin2000,guo19}, photon-echo using atomic frequency comb (AFC)~\cite{afzelius2009,afzelius2010,afc2010,jobez2016} and intra-atomic frequency comb (I-AFC)~\cite{iafc2019,teja2021,iafc2021}.  All these techniques use a large ensemble of atoms or bulk materials to store  photons.

To gain scalability and the practical advantage in quantum information processing, many efforts are being devoted towards integrated photonic chips~\cite{crespi2011,kohnen2011,meng2015,hacker2016,freer2017,titchener2018,elshaari2020,wang2020}.  On-chip single photon sources, on-chip beamsplitters and on-chip photon detectors are already available on integrated platform~\cite{uppu2020,laucht2012,lu2021,yin2022,gyger2021}, while on-chip quantum memory is still a work in progress and is highly sought after device~\cite{zhong2017,liu2020}. Atomic ensemble based quantum memories pose challenges for their on-chip integration.  So far only the  AFC based quantum memory protocol has been repurposed for the on-chip implementation~\cite{zhong2017,liu2020}. Further, the Raman quantum memory protocol has been extended to a single-atom level which can potentially be used on integrated photonic chips~\cite{specht2011}.

In this article, we propose a scheme for storing weak light pulses and single photons using a single atom coupled to an optical cavity. The trapped atom contains an I-AFC. We show that this joint single-atom-cavity setup results in a photon-echo, similar to the I-AFC based quantum memory protocol~\cite{iafc2019}.  The efficiency of storing the light in this setup depends on the finesse of the optical cavity and the quality of the frequency comb. We can also achieve robust and efficient storage for polarization and time-bin qubits using this setup. As examples, we show that Cesium and Rubidium atoms coupled to nanophotonic waveguide cavities can serve as promising candidates for the implementation of this quantum memory protocol.

In principle, the efficiency of this protocol can reach up to $100\%$, whereas in AFC and I-AFC protocols the maximum efficiency in the forward propagation can only be $54\%$. This is because, unlike in the bulk AFC and I-AFC protocols, the reabsorption of the photon in the remission process is eliminated by keeping the atom-cavity setup in the Purcell regime and using only one atom for storing the light.

One of the biggest advantages of the proposed scheme is that it provides a possible realization of an on-chip quantum memory. Furthermore, since our protocol requires only a frequency comb coupled to a cavity, it can also be implemented using the quantum dots inside a cavity~\cite{kiraz2003,an2007,zhang2018,westmoreland2019}. On-demand single-photon sources have already been realized using quantum dots~\cite{heinze2015,liu2018,uppu2020,lu2021}. Combining these two can pave the way for efficient on-chip photonic quantum computation.

The article is organized as follows: In Sec.~\ref{Sec:Background}  we introduce the relevant background required for our results. In Sec.~\ref{Sec:Results}, we present our result where we show the photonic quantum memory using a single atom trapped inside an optical cavity. The examples of Rubidium and Cesium atoms for the implementation of the quantum memory are presented in this section. We conclude in Sec.~\ref{Sec:Conclusion}.

\section{Background}\label{Sec:Background}
In this section,  we introduce the I-AFC based quantum memory protocol and the dynamics of an atom-cavity setup interacting with electromagnetic field.

\subsection{Intra-atomic frequency comb}

In I-AFC based quantum memory, we consider an atom with degenerate ground and excited hyperfine levels. This degeneracy is lifted by applying an external magnetic field which results in multiple ground and excited states.  All these  multiple dipole allowed transitions considered together result in a comb like structure known as I-AFC~\cite{iafc2019}.

The interaction picture Hamiltonian for an atom that exhibits I-AFC, interacting with electric field $\mathcal{E}(z,t)$ with mean frequency $\omega_L$ reads 
\begin{align}
H=\hbar  \sum_{n=1}^{N} \delta_n \dyad{e_n}{e_n} 
-\hbar \qty[\sum_{n} \Omega_{n}(z,t) \dyad{e_n}{g_n}+\text{h.c.}].
\end{align}
We have considered the atom with $N$ number of ground states $\{\ket{g_n}\}$ and $N$ number of excited states $\{\ket{e_n}\}$. For simplicity,  the  transition is allowed only between $\ket{e_n}\leftrightarrow\ket{g_n}$ for all $n$ and the corresponding transition dipole moment is given by $d_{n}$. Here, the Rabi frequency $\Omega_{n}(z,t)={d_{n}\mathcal{E}(z,t)}/{2\hbar}$, and $\delta_{n}=\omega_{n}^e-\omega_{n}^g-\omega_L$ is the detuning between the $\ket{e_n}\leftrightarrow\ket{g_n}$ transition frequency and $\omega_L$.

 All these dipole allowed multiple transitions collectively yield a frequency comb. The line width of each of the transition is $\gamma_n$ and the mean frequency is $\delta_n + \omega_L$. The spacing between the $(n+1)$-th and the $n$-th teeth is  given by $\Delta_n = \delta_{n+1} -\delta_n$. For simplicity, we consider an ideal I-AFC with uniform comb spacing $\Delta_n \equiv \Delta$ and equal tooth width $\gamma$ such that the detuning can be written as $\delta_{n}\equiv n \Delta$.

 We consider the absorption of a single photon pulse $\mathcal{E}(0,t)$ with spectral width $\gamma_p$ in the I-AFC such that $\gamma_p\gg\Delta$. If the atom is initially prepared in an equal superposition of multiple ground states, i.e., $\ket{G}=\dfrac{1}{\sqrt{N}}\sum_n{\ket{g_n}}$, the initial state for the ensemble of $M$ atoms becomes  $\ket{G}^{\otimes M}$.
The state of the I-AFC at time $t$ after absorbing the single photon pulse can be written as
\begin{align}
\ket{\psi(t)} \equiv \sum_{j=1}^M \qty(\alpha_j \ket{G}^{\otimes (M-1)}\ket{E(t)}_j), \label{cstate}
\end{align}
where the index $j$ runs over the number of atoms in the ensemble. The state $\ket{E(t)}_j=\dfrac{1}{\sqrt{N}}\sum_n e^{\mi \delta_n t}\ket{e_n}_j$ represents the collective excited state of the $j^{th}$ atom and $\alpha_j$ denote the absorption coefficient of each atom.

The probability of the photon-emission is given by $P(t) \propto \abs{\mel {G}{\sum_jS_j}{\psi}}^2$, where $S_j=\sum_{n}{\ket{g_n}_{j}{\bra{e_n}_j}}$ is the step down operator for the $j$-th atom~\cite{iafc2019}. Using $\delta_{n}= n \Delta$ in Eq.~\eqref{cstate}, we see that the probability of the photon emission maximizes at $t=2\pi m/\Delta$ for integer m. The output light corresponding to different m values are called photon echoes. The time of echo can be adjusted by changing $\Delta$ which in turn is controlled by the strength of the applied magnetic field.

To achieve on-demand storage using this protocol, the excitation is transferred from the excited level to a spin state with long lifetime by applying a $\pi$-pulse. Applying another $\pi$-pulse transfers the excitation back to the excited state causing the photon-echo at an appropriate time.

The efficiency $\eta$ of the I-AFC quantum memory protocol is defined as the ratio of the intensity of light obtained in the first echo to the total intensity of the input light, which reads as~\cite{iafc2019}
\begin{align}
\eta= \dfrac{\int_{\pi/\Delta}^{3\pi/\Delta} \abs{\mathcal{E}(z=L,t)}^2 \, dt}{\int \abs{\mathcal{E}(z=0,t)}^2 \, dt}. \label{eta}
\end{align}
where $L$ is the length of the atomic ensemble along the direction of propagation of light.
The maximum efficiency that can be achieved using the standard I-AFC scheme is $54\%$ in forward mode and $100\%$ in backward mode~\cite{iafc2019}.

\subsection{ Dynamics of an atom-cavity system interacting with electromagnetic field}\label{acs}

Consider a two-level atom coupled to an optical cavity. Let $\{\ket{g},\ket{e}\}$ be the two energy levels of the atom, $\hat a$ be the annihilation operator of the cavity mode, $\omega_c$ and $\omega_{eg}$ be the cavity mode frequency and the atomic transition frequency, respectively. The Hamiltonian for the atom-cavity setup can be written as \cite{reiserer2015}
\begin{align}
H&= \hbar\omega_c{\hat a}^\dagger \hat{a} + \hbar \omega_{eg}\dyad{e}{e}
-\hbar \qty[g \sigma_{eg}\hat{a}+ g^*\sigma_{ge} \hat{a}^\dagger],
\end{align}
where $\sigma_{ij} \equiv \ket{i}\bra{j}$ represents the transition operators for the two-level atom, $g=\dfrac{d_{eg}}{\hbar}\sqrt{\dfrac{\hbar \omega_c}{2\epsilon_0V}}$ is the coupling constant between the atom and the cavity, and $d_{eg}$ is the transition dipole moment between the levels $\ket{e}$ and $\ket{g}$.

The optical cavity is also coupled to an input electromagnetic field mode $\hat a_{\rm in}$ and yields an output mode $\hat a_{\rm out}$ upon interaction of the input field with the cavity. Using the standard input-output formalism~\cite{gardiner1985}, the input, output and the cavity modes are related as 
\begin{align}
  {\hat{a}}_{\text{out}}(t)- {\hat{a}}_{\text{in}}(t)&=\sqrt{\kappa}~ {\hat{a}}(t).\label{tin}
\end{align}
Our goal is to calculate the output field mode $\hat a_{\rm out}(t)$ as a function of time for the given input field mode $\hat a_{\rm in}(t)$. In order to achieve that we need to solve for $\hat a(t)$. The dynamical equations for the atom-cavity setup can be written as~\cite{gardiner1985}
\begin{align}
\dv{}{t} \hat{a}(t) &= \dfrac{1}{\mi \hbar}[\hat a, H]  -\qty(\dfrac{\kappa}{2}) \hat{a} -\sqrt{\kappa} \hat{a}_{\text{in}},\label{ip1}  \\
\dv{}{t} {\sigma_{ge}}(t) &=  \dfrac{1}{\mi \hbar}[\sigma_{ge}, H]-\qty(\dfrac{\gamma}{2})\sigma_{ge}, \label{ip2}
\end{align}
where $\kappa$ is the decay rate of the cavity mode at which it leaks out from the cavity and $\gamma$ is the free space spontaneous emission rate of the atom.

The atom-cavity setup  is usually operated in two parameter regimes, (i) the strong coupling regime where the atom-cavity coupling is the highest i.e. $g \gg \kappa,\gamma$.  As a result, the system undergoes a series of damped Rabi oscillations before the photon leaks out of the cavity~\cite{reiserer2015}. (ii) The bad cavity regime or Purcell regime in which the cavity decay rate $\kappa$ is maximum and is characterized by \mbox{$\kappa>g^2/\kappa>\gamma$}~\cite{agarwal2012}.
In the Purcell regime, the atomic decay rate into the cavity mode is enhanced over the free space decay, as a result of which the atom predominantly decays into the cavity mode~\cite{agarwal2012}. We use this regime in our protocol, since it enables the emission of the photon-echo into the cavity mode, and the emitted photon quickly leaks out of the cavity at a rate $\kappa$ before it can get reabsorbed.

\section{Results}\label{Sec:Results}

 The conventional I-AFC based quantum memory scheme uses an ensemble of atoms to store single photons. However, since each atom in I-AFC contains a frequency comb, it can be argued that a single atom is capable of storing photons provided the atom and light couple strongly. In principle, one can use a single mode optical cavity to tune the coupling between the atom and the photons and can realize quantum memory using a single atom. In this section, we explore this feasibility and show that with a proper choice of atom-cavity parameters, one can realize an efficient quantum memory.  We also present the implementation scheme using Rubidium and Cesium atoms, as examples, coupled to  nanophotonic waveguide cavities.

\subsection{Quantum memory using single-atom I-AFC coupled to a cavity} \label{saqm}
 
\begin{figure}[!htb]
\subfigure[\label{iafccav}]{\includegraphics[width=6cm]{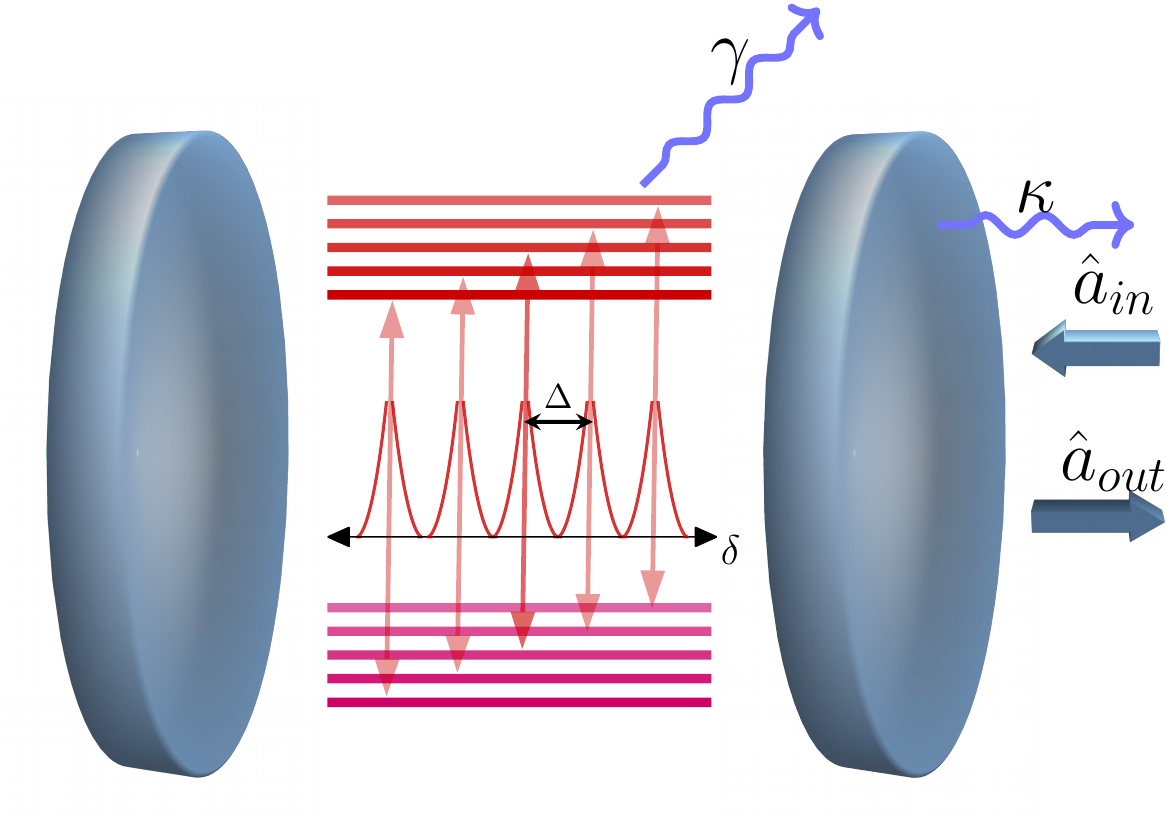}}
\hspace{1cm}
\subfigure[\label{figa}]{\includegraphics[width=6cm]{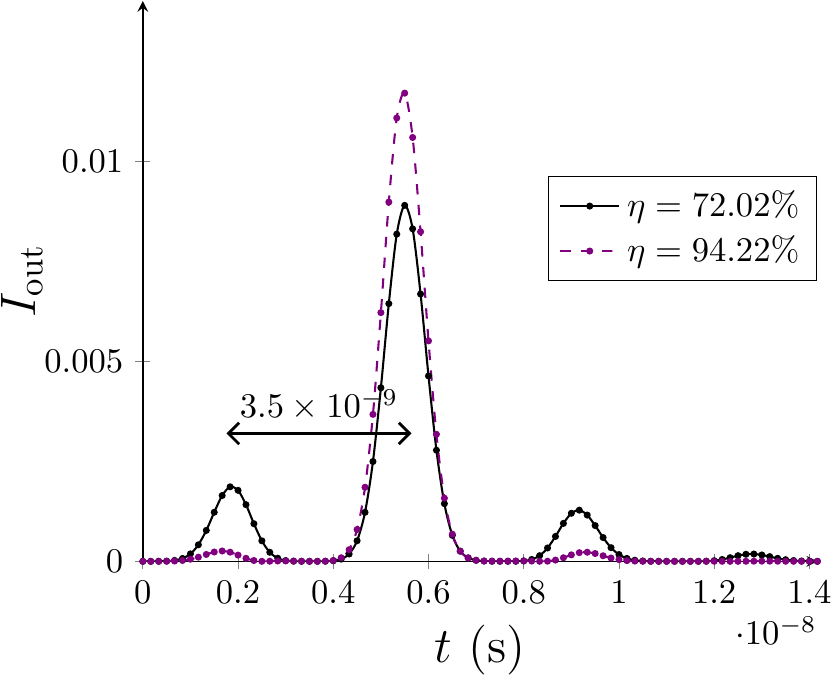}}
\caption{\subref{iafccav} Schematic diagram for an I-AFC inside a cavity. Here, the I-AFC is interacting with a single cavity mode with decay rate $\kappa$. $\hat{a}_{\rm in}$ and $\hat{a}_{\rm out}$ represent the input and output cavity field operators. $\gamma$ is the spontaneous decay rate of the atom into free space. \subref{figa} Photon-echo after a delay of $3.5$ns for an ideal IAFC coupled to a cavity with uniform comb spacing of $\Delta=300$ MHz, tooth width $\gamma = 7.5$ MHz and cavity detuning $\Delta_c=0$. The two photon echoes shown in dashed and solid curve correspond to the cavity decay rate $7$ and $4$ GHz, respectively, with the corresponding efficiencies $94.22\%$ and $72.02\%$, respectively.}\label{f1a}
\end{figure}

Consider an atom that contains a frequency comb, coupled to a high finesse single-mode optical cavity~ [Fig.~\ref{iafccav}].
The Hamiltonian for such atom-cavity system consists of three parts, the free Hamiltonian of the single-mode cavity, the free Hamiltonian of the atom and the interaction between the two systems, which reads
\begin{align}\label{ham}
H=&H_{\text{c}}+H_{\text{a}}+H_{\text{int}}\nonumber\\
\begin{split}
=&\hbar \omega_c\hat{a}^\dagger \hat{a} + \sum_{n=1}^{N_e}\hbar \omega_n^e\dyad{e_n}{e_n}+\sum_{m=1}^{N_g}\hbar \omega_m^g\dyad{g_m}{g_m}\\
&-\hbar\qty[\sum_{n,m} g_{nm}\dyad{e_n}{g_m}\hat{a}+\sum_{n,m} g_{nm}^*\dyad{g_m}{e_n} \hat{a}^\dagger],
\end{split}
\end{align}
where $\hat{a}$ is the photon annihilation operator for the cavity mode.
$\ket{g_m}$ and $\ket{e_n}$ denote the $m$-th ground state and the $n$-th excited state, respectively, with coupling strength 
$g_{nm}=\dfrac{d_{nm}}{\hbar}\sqrt{\dfrac{\hbar \omega_c}{2\epsilon_0V}}$. The $d_{nm}$ is the transition dipole moment between $\ket{g_m}\leftrightarrow\ket{e_n}$ transition and  $\omega_c$ is the resonance frequency of the cavity.

 The dynamical equations for the cavity field operator $\hat{a}$ and the atomic lowering operator $\sigma_{mn}^-\equiv\dyad{g_m}{e_n}$ using input-output formalism as discussed in Sec.~\ref{acs} read
\begin{align}
\dv{\hat{a}}{t}&=-i\omega_c\hat{a}+i\sum_{n,m}g_{nm}^*\sigma_{mn}^{-}-\dfrac{\kappa}{2}\hat{a}-\sqrt{\kappa}{\hat{a}_{\text{in}}}, \label{r1} \\ 
\dv{{\sigma}_{mn}^-}{t}&=-\mi(\omega_n^e-\omega_m^g)\sigma_{mn}^-+\mi g_{nm}(\sigma_{mm}-\sigma_{nn})\hat{a} -\dfrac{\gamma}{2}\sigma_{mn}^-, \label{r2} \\   
\sqrt{\kappa}~\hat{a}(t) & = \hat{a}_{\text{out}}(t)-\hat{a}_{\text{in}}(t). \label{r3}
\end{align}
Here $\gamma$ is the spontaneous decay rate of the atom in free space and $\kappa$ is the decay rate of the cavity field.

Solving these equations in the frequency domain using the low intensity approximation ($\expval{\hat{a}^\dag{_\text{in}} \hat{a}_{\text{in}}}\lesssim 1$) which amounts to $\sigma_{nn} \thickapprox0$~\cite{hu2015} yields
\begin{align}
{\hat{a}}_{\text{out}}(\omega) =\qty[1-\dfrac{\kappa}{\mi(\omega+\Delta_c)+\mathcal{D}(\omega)+\dfrac{\kappa}{2}}]{\hat{a}}_{\text{in}} (\omega). \label{e6}
\end{align}
Here
\begin{align}
\mathcal{D}(\omega)&=\sum_{n,m}\dfrac{\sigma_{mm} \abs{g_{nm}}^2}{\qty[i(\omega+\delta_{nm})+\dfrac{\gamma}{2}]}, \label{e7}
\end{align}
is the I-AFC propagator~\cite{iafc2019}, and $\Delta_c=\omega_c-\omega_L$, $\delta_{nm}=(\omega_n^e-\omega_m^g)-\omega_L$ are the detunings with respect to the input light.  
Inverse Fourier-transform of Eq.~(\ref{e6}) yields the output field in time $\hat a_{\rm out}(t)$~\cite{milburn2015}. 

In order for this atom-cavity setup to qualify for a quantum memory, there must be a delay between the input and output light. We solve Eq.~(\ref{e6}) numerically taking the initial state of the atom as an equal superposition of ground states and plot the intensity of the output field $I_{\rm out}=\left\langle \hat a^\dagger_{\rm out}(t)\hat{a}_{\rm out}(t)\right\rangle$ as function of time $t$ [Fig.~\ref{figa}]. Here we have considered an atom with an I-AFC having seven teeth with uniform comb spacing  $\Delta=300$ MHz, tooth width $\gamma = 7.5$ MHz and detuning  $\Delta_c=0$. The solid curve and the dashed curve in this figure corresponds to different cavity decay rates $\kappa$. In this figure, we can clearly see that the first prominent output pulse of light is at time $t = 2\times 10^{-9}$s which is due to the immediate reflection from the cavity. The second  prominent output pulse occurs at $t \sim 5.5\times 10^{-9}$s which is due to the emission from the cavity. There is a delay of $3.5$ns which is approximately  $2\pi/\Delta$ due to the interaction of light with the setup. Hence the atom-cavity setup behaves like an I-AFC.

\begin{figure*}[!htb]
\subfigure[\label{figb}]{\includegraphics[width=4cm]{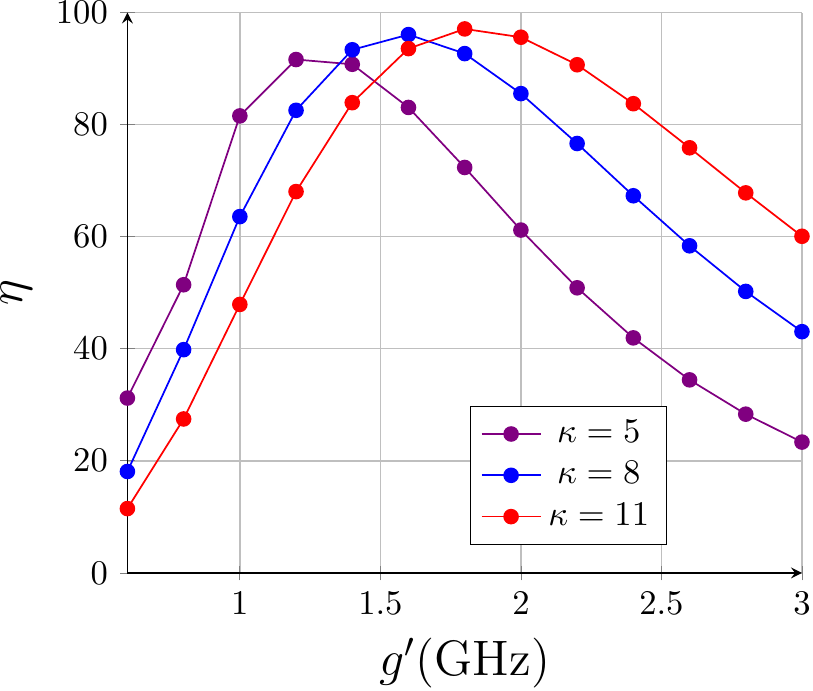}}
\subfigure[\label{figb1}]{\includegraphics[width=4cm]{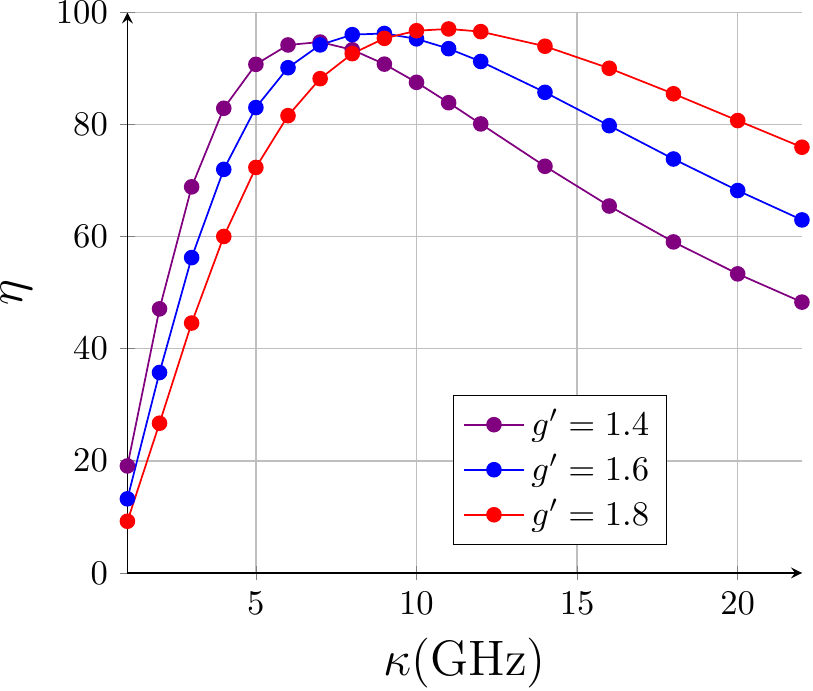}}
\subfigure[\label{figb2}]{\includegraphics[width=4cm]{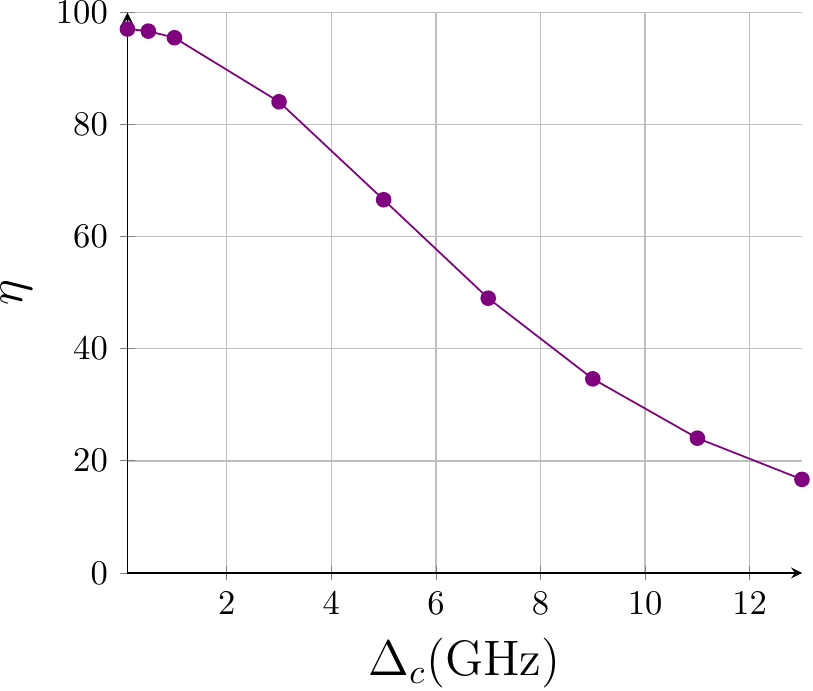}}
\subfigure[\label{finesse}]{\includegraphics[width=4cm]{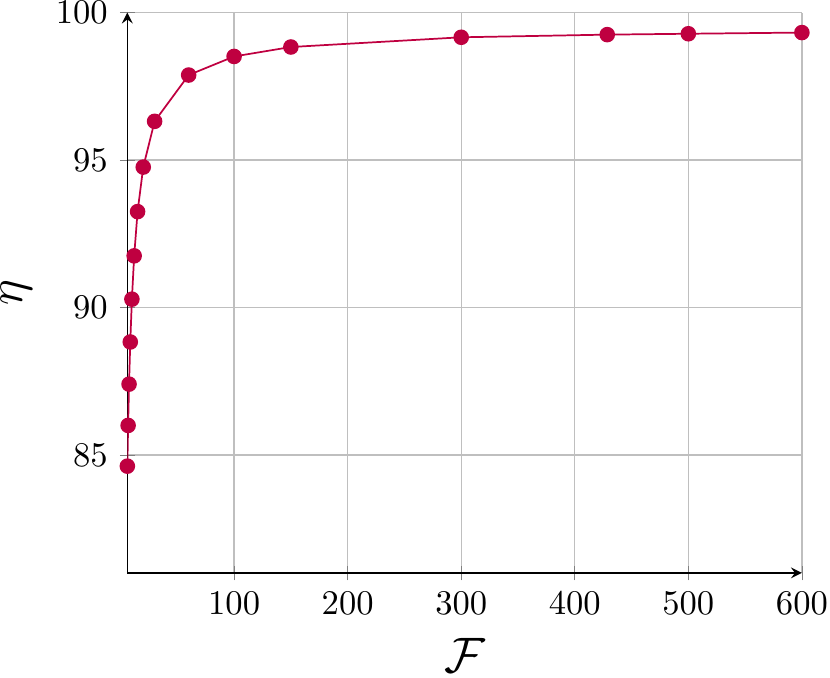}}
\caption{Effect of various parameters on the efficiency ($\eta)$ of quantum memory in the single-atom-cavity setup for an ideal comb with uniform comb spacing $\Delta=300$ MHz.
In \subref{figb} we plot the variation of $\eta$ as a function of $g'$ for fixed values of $\kappa$. \subref{figb1} shows the variation of $\eta$ as a function of $\kappa$ for fixed values of $g'$. In \subref{figb2} and \subref{finesse}, we plot the variation of $\eta$ as a function of the cavity detuning $\Delta_c=\omega_c-\omega_L$ and the comb finesse ($\mathcal{F}$), respectively for $(g',\kappa)=(1.8,11)$ GHz.}\label{f1}
\end{figure*}

Eqs.~\eqref{e6} and~\eqref{e7} suggest that the cavity parameters $g_{nm},\kappa$ and $\Delta_c$ also play a role in the output field and can affect the quality of the memory. To quantify the quality of the quantum memory we can generalize the definition of the efficiency [Eq.~\eqref{eta}] for the bulk I-AFC protocol to the current scenario as 
\begin{align}
\eta= \dfrac{\int_{\pi/\Delta}^{3\pi/\Delta} \expval{\hat a^\dagger_{\text{out}}(t) \hat a_{\text{out}}(t)} \, dt}{\int \expval{\hat a^\dagger_{\text{in}}(t) \hat a_{\text{in}}(t)} dt}. \label{etac}
\end{align}

For an ideal I-AFC, since all the peaks are identical, i.e., $d_{nm} \equiv d$, we may write $g_{nm}=g$. We define  {$g'=\sqrt{\sigma_{mm}} g$}  as the effective coupling constant. In Fig.~\ref{figb}, we plot the variation of $\eta$ as a function of $g'$ keeping $\kappa$ constant and cavity detuning $\Delta_c = 0$. We have considered the I-AFC having seven teeth with uniform comb spacing $\Delta = 300$ MHz and tooth width $\gamma = 7.5$ MHz. We have also numerically optimized the efficiency with respect to the spectral width of the incoming pulse for each parameter. Fig.~\ref{figb} shows that an optimum value of $g'$  exists
for every given value of $\kappa$ which maximizes the efficiency.
A similar trend is observed, when we vary $\kappa$ with fixed value of $g'$ in Fig.~\ref{figb1}. 
From these two figures, we can find that the efficiency is maximum for $(g',\kappa) = (1.8,11)$ GHz for our case.

In Fig.~\ref{figb2}, we plot the variation of the efficiency $\eta$ as a function of the cavity detuning, $\Delta_c(=\omega_c-\omega_L)$ while keeping $(g',\kappa) = (1.8,11)$ GHz. As expected, it shows a drop in the efficiency as the cavity detuning $\Delta_c$ increases.
Apart from these parameters, the optimized efficiency also depends on the comb finesse $\mathcal{F}$ which is defined as the ratio of the comb spacing and the peak width i.e. $\mathcal{F}\equiv\Delta/\gamma$. In  Fig.~\ref{finesse}, we plot the efficiency as a function of comb finesse for the ideal comb with fixed comb spacing, $\Delta=300$ MHz by changing the peak width $\gamma$ while keeping the cavity parameters to be fixed at $(g',\kappa)=(1.8,11)$ GHz. This plot shows that the efficiency saturates to $\sim 100\%$ asymptotically for high values of finesse.

Note that the solution for the output field in Eq.~\eqref{e6} is derived under the approximation that there is negligible absorption of the input field by the atom ($\sigma_{nn}\sim 0$), however the effficiency of the storage by this system is still high. To understand this, we consider the expression for the susceptibility $\chi$ of the joint atom-cavity system, which reads~\cite{chang2011}
\begin{align}
\chi =\sqrt{\dfrac{2  V}{\epsilon_0\hbar \omega_c}}\dfrac{\expval{P}}{ {\expval{a}}},\label{chi}
\end{align}
where $P=\dfrac{1}{V}\sum_{n,m}d^*_{mn}\sigma_{mn}^-$ is the atomic polarization of the  atom exhibiting the I-AFC and $V$ is the cavity mode volume. 
The above expression for susceptibility is equivalent to the classical field susceptibility $\chi_e=\dfrac{\expval{P}}{\epsilon_0 \mathcal{E}}$, where the classical field amplitude $\mathcal{E}$ being replaced by the expectation value of $\hat{a}$ operator.

\begin{figure}[!htb]
\includegraphics[height=4.5cm,width=6cm]{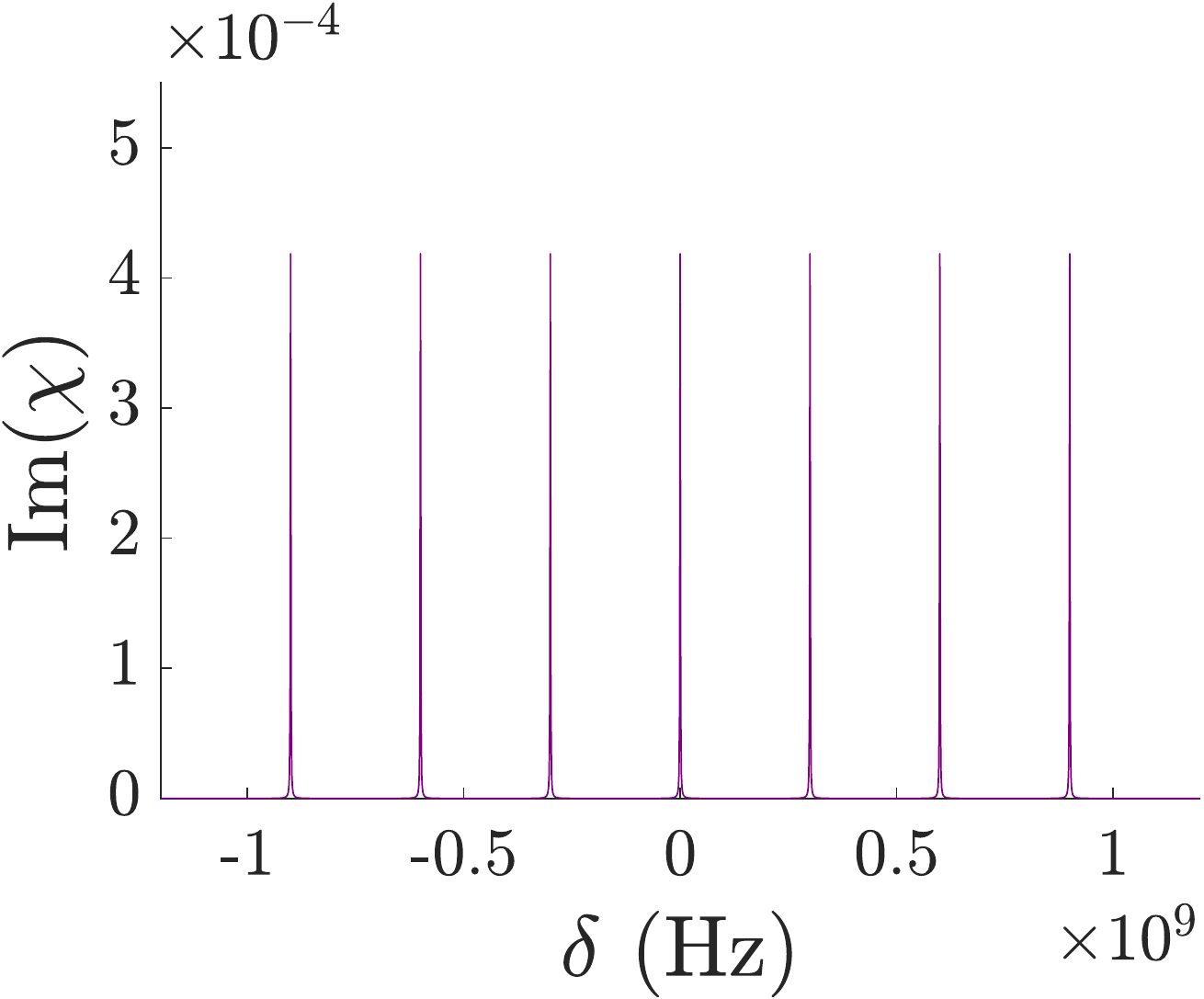}
\caption{Plot for absorption of joint atom-cavity system for an ideal frequency comb with uniform comb spacing $\Delta=300$ MHz, tooth width $\gamma = 7.5$ MHz and cavity detuning $\Delta_c=0$.}
\label{comb}
\end{figure}

In Fig.~\ref{comb} we plot the absorption for the atom-cavity system where the absorption is the imaginary part of the joint susceptibility $\chi$. From this figure,  we can see that the absorption profile of the joint atom-cavity system  shows the comb like structure  similar to the absorption profile in I-AFC. This comb like structure is responsible for the photon-echo, as shown in Fig.~\ref{figa}. Thus the atom and the cavity together account for the photon storage, even though the absorption by the atom is negligible.

\begin{figure*}[!thb]
\subfigure[\label{figrb}]{\includegraphics[height=5cm,width=5.5cm]{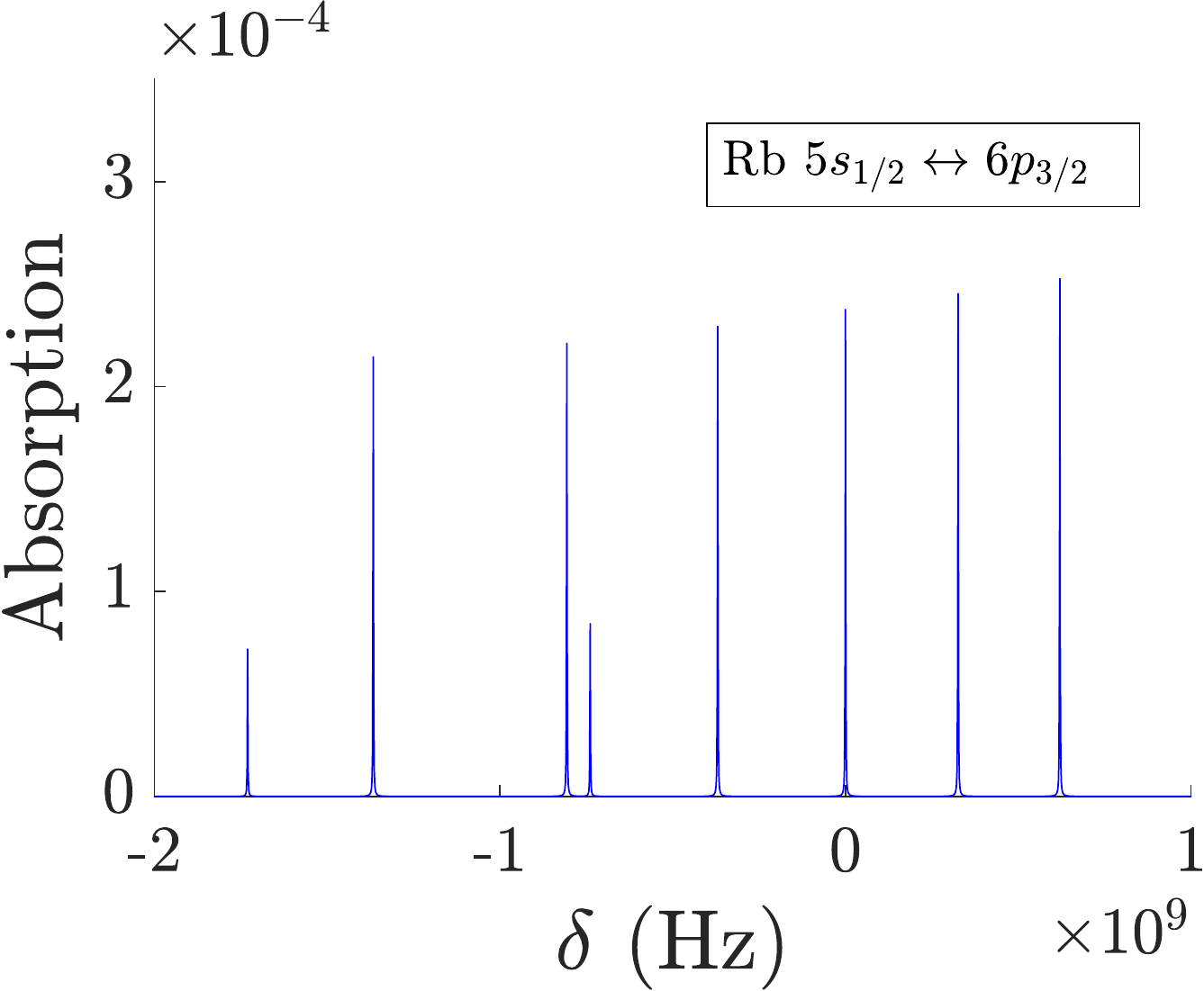}}
\hspace{2.5pt}
\subfigure[\label{figcs}]{\includegraphics[height=5cm,width=5.5cm]{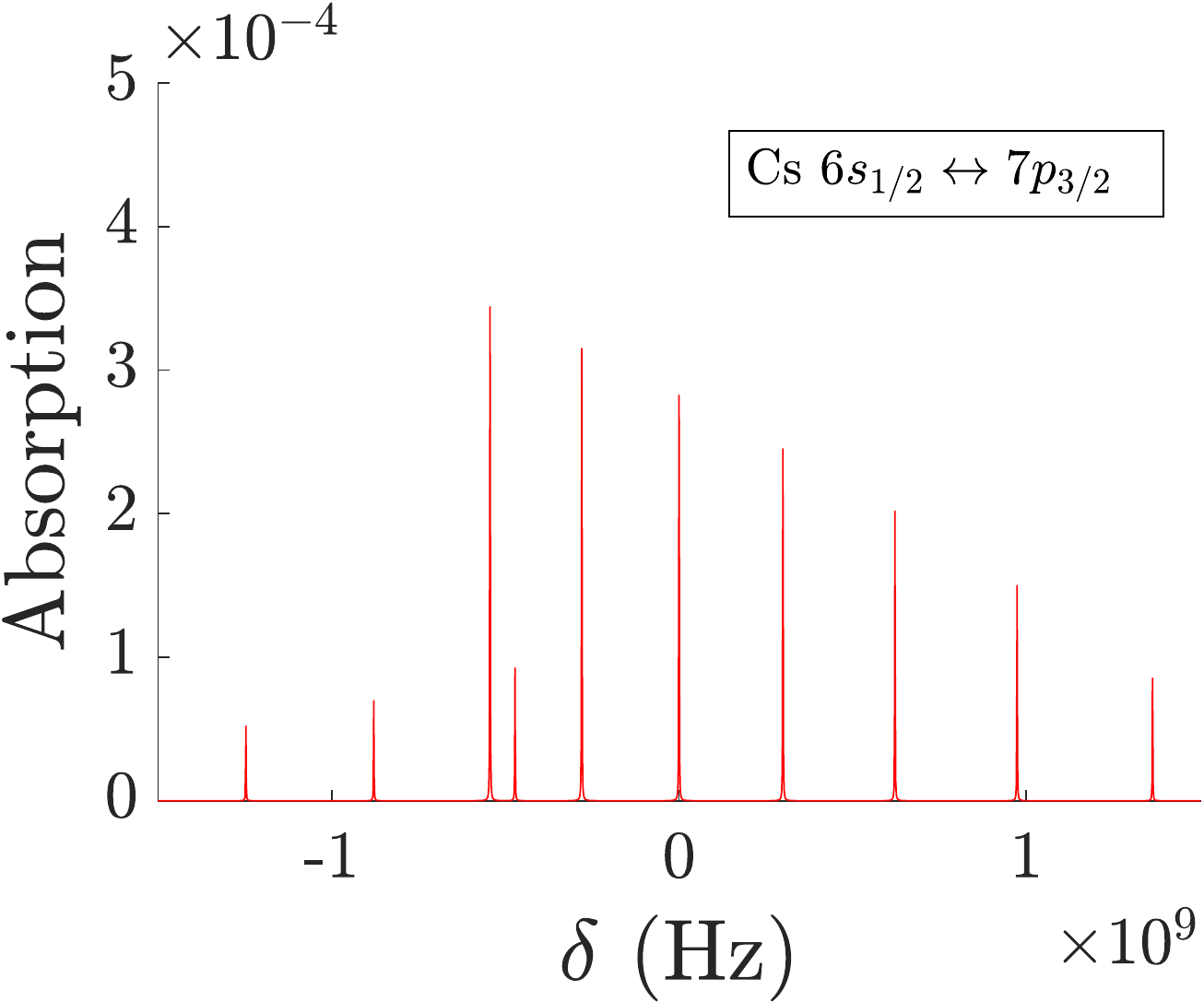}}
\caption{\subref{figrb} and \subref{figcs} represent frequency comb in Rb and Cs atoms for transitions between $5s_{1/2}\leftrightarrow 6p_{3/2}$ for Rb and $6s_{1/2} \leftrightarrow 7p_{3/2}$ for Cs. The applied magnetic field strength for Rb and Cs are taken to be 0.15 and 0.1 T, respectively.}\label{acomb}
\end{figure*}

This scheme can also be used to store polarization and time-bin photonic qubits. AFC and I-AFC based quantum memories are known for storing time-bin qubits efficiently~\cite{afzelius2009,gundougan2015,ortu2022}. Moreover, it has been shown that I-AFC can store polarization states of light \cite{iafc2021}. For storing polarization qubit, we can consider the same atom-cavity setup with the single atom consisting of two overlapping frequency combs corresponding to two different polarizations. Generally, the efficiency and the photon-echo time for the two polarizations can be different. By choosing the cavity parameters appropriately, one can store arbitrary polarization in these systems~\cite{iafc2021}.

To conclude this section, we have shown that a single atom with I-AFC coupled to an optical cavity can store photons efficiently. We have shown the effect of various parameters on the quality of the storage and estimated the optimum values of the  parameters for the most efficient storage. The results obtained here are also interesting from a fundamental point of view. We see that even though the I-AFC is necessary to store the photons in the atom-cavity system proposed here, the excitation probability of the atom is negligible. The interaction of the I-AFC with the cavity yields the comb like absorption profile of the joint atom-cavity system, which enables efficient quantum memory. In the following section, we present  examples of systems capable of realizing this quantum memory protocol. 

\subsection{Realizing the quantum memory using Rb and Cs atoms} \label{Implemen}

\begin{figure*}[!thb]
\subfigure[\label{figc}]{\includegraphics[height=4.8cm,width=6.5cm]{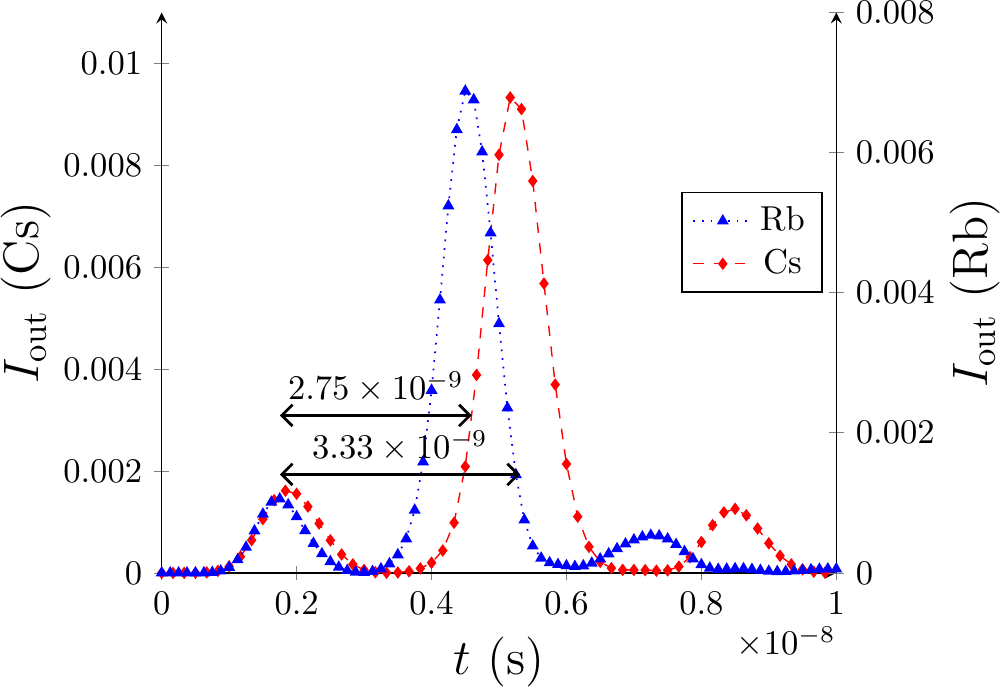}}
\hspace{2.5pt}
\subfigure[\label{figv}]{\includegraphics[height=4.8cm,width=5cm]{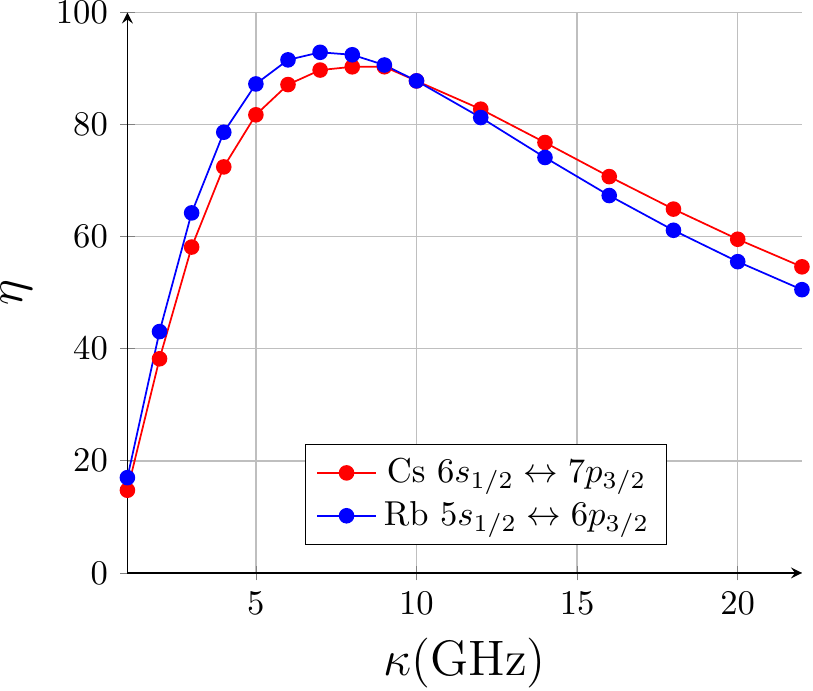}}
\caption{\subref{figc} Photon-echo for the IAFC in Rb and Cs atoms. \subref{figv} Variation of efficiency as a function of $\kappa$ in Rb and Cs atoms.}\label{eff}
\end{figure*}

So far, we have discussed the photon storage assuming an ideal frequency comb with uniform comb spacing and having same peak height, which interacts with an optical cavity. However, if we consider realistic systems such as Rb and Cs atoms, the frequency combs obtained from them are usually non-uniform with unequal peak heights which affects the storage process~\cite{iafc2019,teja2021}. In this section, we discuss the possibilities for experimental implementation of the single atom based quantum memory protocol in realistic systems such as Rb and Cs atoms coupled to nanophotonic waveguide cavity and show that the current scheme can be implemented with the existing experimental techniques.

As discussed in Sec.~\ref{saqm}, one of the requirements to achieve efficient quantum memory in I-AFC-cavity setup is a cavity with high coupling strength $g$ of the order of GHz (see Fig.~\ref{f1}). This, in turn requires cavity with low mode volume of the order of $(\sim \mu \text{m})^3$. Such strong coupling is difficult to achieve using the conventional Fabry-P\'erot cavities, but can be achieved using the nano-cavities~\cite{van2011,thompson2013} where mode volume $V \sim \lambda^3$ have already been realized. 
Apart from this, the strong coupling has been realized in fiber-based Fabry-P\'erot cavity ~\cite{hunger2010} where the mirror surface of the cavity is designed on the optical fiber end faces. This tight confinement using nano-photonic cavities gives an additional advantage of potential integration with nano-photonics. Trapping in such low mode volumes results in the atom-cavity strong coupling of the order of $g\sim$ GHz along with the quality factor $Q=\omega_c/\kappa \sim 10^5$~\cite{van2011}.

\begin{table}[!htb]
\small
\begin{tabular}{|c|c|c|c|c|c|c|}
\hline 
Atom & Transition  & $\lambda$ (nm) & $B$ (T) & $V(\mu \text{m})^3$ & $\kappa$ (GHz) & $Q$\\ 
\hline 
Rb~\cite{sansonetti2006}& $5s_{1/2}\leftrightarrow 6p_{3/2} $ &420.3 & 0.15 & $20$ & $\sim 7$ & $10^5$\\ 
\hline 
Cs~\cite{sansonetti2009}& $6s_{1/2} \leftrightarrow 7p_{3/2} $ &455.66 & 0.1 & $ 20$ & $\sim 8$ & $10^5$\\  
\hline
\end{tabular} 
\caption{Rb and Cs parameters used in numerical calculations. $\lambda$ is the wavelength of transition $B$ is the magnetic field used in obtaining IAFC. $V, ~\kappa$ and $Q$ are the mode volume, decay rate and quality factor of the cavity, respectively.}\label{tab}
\end{table}

Although the scheme presented in this paper is applicable to a large class of atoms, molecules and quantum dots, here we consider Cs and Rb atoms as examples to realize this quantum memory protocol. The parameters such as the transitions, the wavelength, applied magnetic field strength and so on for Rb and Cs atoms used for our calculations are given in Table~\ref{tab}. 
In Figs.~\ref{figrb} and \ref{figcs}, we plot the frequency comb obtained in  Rb and Cs atoms. Clearly, these frequency combs are neither uniform in the comb spacing nor do they have equal absorption peaks. In Fig.~\ref{figc} we show the photon-echo from Rb and Cs atoms  calculated numerically by solving Eq.~\eqref{e6}. The maximum efficiencies for Rb and Cs atoms are found to be $92.9\%$ and $90.36\%$, respectively, for the parameters specified in Table.~\ref{tab}.

In Fig.~\ref{figv}, we plot the variation of the efficiency as a function of the cavity decay rate $\kappa$ for Rb and Cs atoms. It is clear that the trend is similar to that of an ideal comb with peak value of $\sim 90\%$. The lesser value of the efficiencies in the case of Rb and Cs atoms is due to the inherent non-uniformity present in the frequency combs. This non-uniformity is attributed to different values of the comb spacing $\Delta_{n}$ and the dipole matrix element $d_{nm}$ corresponding to the transition $\ket{e_n}\leftrightarrow \ket{g_m}$. Our calculations show that an efficient quantum memory using a single atom coupled to an optical cavity can be implemented using  the current experimental techniques.

\section{Conclusion}\label{Sec:Conclusion}

On-chip photonic quantum memories are essential for scalable and integrated photonic quantum information processing. Most of the currently available quantum memory protocols require atomic ensembles or bulk materials to store photons. Here we have proposed a new scheme to store photons using only a single atom coupled to an optical cavity. The atom exhibits an I-AFC which enables the joint atom-cavity  system to store photons.  This provides us with a possibility to realize an on-chip quantum memory suitable for integrated photonic chips. The proposed setup is capable of storing time-multiplexed photons, along with their polarization degree of freedom efficiently, hence providing multi-mode photonic quantum memory. Theoretically, our quantum memory protocol can store photons with $\sim100\%$ efficiency. Although we have presented this quantum memory protocol using trapped atoms, this can very well work with quantum dots and quantum defect centers. The advantage of working with the quantum dots and the defect centers is that for the case of atoms, the working temperature of the protocol is $10^{-3}-10^{-6}$ K, whereas this temperature is $\sim 1$ K for quantum dots. Since deterministic single photon sources have already been realized using quantum dots, combining it with the on-chip quantum memory can provide a robust integrated platform for photonic quantum computation.

\begin{acknowledgments}   
  Chanchal acknowledges the Council of Scientific and Industrial Research (CSIR), Government of India, for financial support through a research fellowship (Award No. 09/947(0106)/2019-EMR-I). S.K.G. acknowledges the financial support from the Inter-disciplinary Cyber Physical Systems (ICPS) program of the Department of Science and Technology, India (Grant No. DST/ICPS/QuST/Theme-1/2019/12).

\end{acknowledgments}


%

\end{document}